\documentclass{iaus260}
\usepackage{graphicx}
\pubyear{2009}
\volume{260}  %% insert here IAU Symposium No.
\pagerange{10--17}
\jname{The R\^ole of Astronomy in Society and Culture}
\editors{D. Valls-Gabaud \& A. Boksenberg, eds.}
\title[Astronomy and the Media]                   %% Give here short title %%
{Astronomy and the Media: a love story?}  %% Give here full title  %%

\author[Henri Boffin]           %% Give here short author list %%
{Henri M.J. Boffin}        %% Give here full author list %%

\affiliation{ESO \\ 
                 Karl-Schwarzschild-str. 2, 85748 Garching, Germany \\ 
                 email: {\tt hboffin@eso.org} 
                 }

\begin{document}
\maketitle

\begin{abstract}
With the availability of nice images and amazing, dramatic stories, the fundamental questions it addresses, and the attraction is exerces on many, it is often assumed that astronomy is an obvious topic for the media. Looking more carefully, however, one realises that the truth is perhaps not as glamorous as one would hope, and that, although well present in the media, astronomy's coverage is rather tiny, and often, limited to the specialised pages or magazines. 
\keywords{science communication, media, astronomical topics, press releases}    %% Add here a maximum of 10 keywords
\end{abstract}

\firstsection % If your document starts with a section,
\section{Introduction}

In astronomy as in other scientific or societal fields, communication is an important aspect that no single organisation can overlook. Especially public research organisations should be accountable to the public for the tax money they use. This is only possible if the public is well informed. But this is even more crucial in order to secure additional funding for new projects. As one scientist said, perhaps a little bit too provocatively, ``{\it the one percent spent on outreach allows one to get the 99 percent to have the project done}''. This is most likely too strong a statement but the general idea is there. Communication is also important to entertain the necessary excellent relations with the local communities -- some of the large astronomical observatories know a lot about this. Communication is also essential for astronomy to fulfil a fundamental role in modern society: attracting bright youngsters to scientific careers. Although girls and boys are more and more moving away from science, there is a great need for future scientists. And even if the young people won't become scientists, it is important that they are sensitive to science as a whole: as grown-ups, they won't be able to avoid relying on science in their daily life, and they will have to take decisions with a scientific dimension. 

For all these reasons, the communication of research organisation will address various target groups: general public, scientists, policy-makers, educators, and the industry. But with limited resources, one needs amplifying outlets to reach a significant fraction of the targeted audience. It is indeed impossible to prepare all types of communication material, with different emphasis, at all levels of complexity, and in all languages. One needs to rely on specific amplifiers. Media outlets are one of these. Indeed, not only are journalists trained to adapt the material to their public, which they know very well, but the public get informed about science through these channels.
The 2007 Eurobarometer on ``Scientific research in the media" (Eurobarometer, 2007) shows for example that 61\% of respondents in the European Union get informed about science watching TV programmes, 49\% reading science articles in general newspapers and magazines, 28\% through the internet, 26\% listening to radio, and 22\% buying specialised press. Similar numbers are observed in the US. Obviously, the media are an important channel to communicate science. However, there are caveats. The first one is that science on TV represents at most 2\% of all news shown. The other is that studies have revealed that only a quarter of all adults can read and understand the stories in the science sections of quality newspapers.

The crucial question is nevertheless whether the media are indeed a efficient channel for communicating astronomy. This is clearly a difficult question which can be answered in several ways. Before briefly attempting to do so, let me make a general remark. As shown above and in various studies, there is no doubt that the media play a a very important role by raising public awareness about science and its results, but it is doubtful how much the media are really able to teach science to the wide public (\cite{West200X}). And one should realise that this is not an easy task. In their study of the public understanding of scientific terms and concepts, the US National Science Foundation (\cite{SciEng2004}) found that less than 15\% of people understand the term ÒmoleculeÓ while less than 50\% know that the Earth goes around the Sun once a year! Starting to talk about gamma-ray bursts, redshifts, galaxies or interferometry represent thus formidable challenges. Scientists and science communicators must thus set realistic goals when interacting with the media and the public, and recognise that other activities are required to transform curiosity into knowledge such as the internet, public events, science centres, and so on. A nice example of such programme, trying to exploit several avenues was the Venus Transit Programme  (Boffin \& West, 2004, 2005). Other examples have been successfully organised in the framework of the International Year of Astronomy 2009. 
 
Coming back to our main question, at first sight there are many reasons to be optimistic and to think that astronomy and the media have a love affair. For example, the American reference newspaper {\it The New York Times} online science section has two specific subsections, one on environment and the other on space \& cosmos! Similarly, the British magazine {\it New Scientist} has a rather successful specific Space section, and one should not forget that the 
{\it BBC Sky at Night} programme is the longest running television series, existing since 1957, although to be fair, one should admit that it is no more shown during prime time but very late in the evening. Here again there is an important caveat, which is that often space and astronomical news are put together, but their share is far from equal. The NSF 2008 study of S\&T attitudes and understanding reveal that the NASA Space Shuttle Programme has taken a very large share of all science related news in 2005 and 2006, but this is of course not astronomy as such.
Another important unfortunate aspect is the general tendency for media to cut down on science coverage. As a journalist of the french newspaper ``Le Monde'' told me, from the 10 journalists working for the science section in 1998, only 4 are still in place ten years later. All others where moved to other sections.

\section{Does astronomy sell?}
In order to try to be a little bit more quantitative, I looked at the US magazine {\it Time}. Since  this main street magazine exists, astronomy has been featured no less than 12 times on its cover. About once every five years or so. This would be a nice result {\it per se}, especially when by comparison, biology had only 4 covers in the same period, and chemistry only 9 (in the latter case, most of them having appeared before 1965). However, when looking at other scientific fields, things start to be less exceptional. History was featured 24 times, and environment took the seat 90 times. The overall winner is definitively medicine which was featured on 248 covers. This is 20 times as much as astronomy! The same trend can be seen in the number of articles dealing with the various topics that appear in the magazine. With 598 articles published from 1923 till nowadays, astronomy comes well behind most other scientific topics. Archeology, biology, chemistry, physics, environment, all do better with, respectively, 1031, 1503, 2240, 2290, and 7764 articles. And again medicine is the great winner with no less than 11814 articles, 20 times as many as the one devoted to astronomy.

This first superficial quantitative study clearly illustrates that while the media do not hesitate to talk about the greatest discoveries in astronomy, it is far from being the most loved of journalists and editors. Is there any logic behind this? Given what I stated above, that journalists know their readers, I would assume so. 
Looking back at an Eurobarometer -- from 2005 this time (\cite{Euro2005}), it is interesting to see that when asked ``which science and technology developments are you most interested in?'', astronomy takes only the 6$^{\rm th}$ place, with 23\% of respondents choosing it. People are more interested in economics and social sciences (24\%), the internet (29\%), humanities (30\%), the environment (47\%), andÉ medicine (61\%). There is clearly a logic, although one could invoke the ubiquitous Òchicken and egg problemÓ as a reason for this situation. Are journalists providing stories on subjects that are most interesting to people or are people interested in the stories given by the journalists? As always, the truth must lie in the middle, but it is perhaps no such surprise that what interests the majority of people is their health. A cause for optimism can be found however in the fact that the comparison between the 2001 and 2005 Eurobarometer surveys reveals an increase of 6\% over 4 years in the percentage of people interested in astronomy. Let us hope that the International Year of Astronomy, with its florilege of activities, will lead to a continuation of this trend. 

%\cite[Anders \& Zinner (1993)]{AndersZinner93} and 
%\cite[Ott (1993)]{Ott93}.
%\cite[Zinner (1998)]{Zinner98}, 

\section{Astronomy topics}

The New York Times science writer John Noble Wilford (as cited by Maran et al., 2000) stated that the topics most likely to cause public impact are mysterious and catastrophic subjects. Astronomy  is not devoid of these and likely candidates would be subjects such as dark matter, black holes, exoplanets, or Near-Earth Objects on collision course with our planets. 

It seems that press offices are not unaware of this and already make a pre-selection along these lines, although some subjects seem more difficult to deal with than others. 
Here is the distribution of subjects in the 144 press releases distributed on the American Astronomical Society (AAS) mailing list, to which about 1300 science journalists are subscribed worldwide, in September and in November 2008: \\
\begin{tabular}{lr}
Solar System &52  \\
New Facilities &15 \\
Exoplanets  &12 \\
Awards, fellowships, contests    &10 \\
Stars, supernovae   &10 \\
Black holes   & 9 \\
Press photos   & 6 \\
Galaxies   & 4 \\
Dark matter  & 2 \\
Cosmology   & 2 \\
\end{tabular}

\vspace{12pt}

Among the press releases distribued by the AAS, one can note the large place taken by the major players. Out of the 144 press releases mentioned above, 61 were issued by NASA (or related to NASA), 17 by ESA and 15 by ESO. The large presence of NASA and ESA could also explain the predominance of solar system stories, as these organisations tend to also devote a large part of their communication to their solar system space missions. But again, things appear more tricky. Looking at  the distribution of topics in all ESO press releases issued between 2004 and 2008 (for a total of 228), one can see that the solar system is also taking its lion's share:\\
  \begin{tabular}{lr} 
 Solar System &52 \\
 Press photos   &43 \\
 Awards, organisation, contests  &42 \\
 Stars, supernovae &38\\ 
 New Facilities   &30 \\
 Exoplanets &20 \\
 Distant Universe & 11\\ 
 Gamma-ray bursts  & 9 \\
 Galaxies   &9 \\
 Black holes  &7\\ 
 Milky Way   &6 \\
 Dark matter  &2\\
  \end{tabular}
\\
ESO being the European intergovernmental organisation for ground-based astronomy, with observatories located in Chile, the space mission argument does not hold here.

\vspace{12pt}

Another way to get into the news (Maran et al. 2000) is to to use a superlative: biggest, most distant, closest, brightest, and so on. Looking at the titles of a few press releases mentioned above show that press officers and scientists are not shy of using these, as shown in the following list:\\
\begin{itemize}
\item  {\bf Closest} Look Ever at the Edge of A Black Hole 
\item  Analysis Begins on Phoenix Lander's {\bf Deepest} Soil Sample 
\item  {\bf First} Picture of Likely Planet Around Sun-Like Star 
\item {\bf Most} Dark Matter-Dominated Galaxy in Universe 
\item  The {\bf Deepest} Ultraviolet Image of The Universe Yet 
\item Gemini Releases Historic Discovery image of Planetary ``{\bf First} Family''
\item  Gamma-Ray Burst was Aimed {\bf Squarely} at Earth.
\end{itemize}

\section{The place of astronomy}
The AAS mailing list is an important source of information for science journalists on astronomy and can clearly serve as a good representation of what science journalists are exposed to. This is particularly relevant as Madsen (2001, 2003) has shown that the astronomy covered in the media finds most of the time its origin in press releases. The first thing to remark is that journalists have a large choice of stories. The AAS distributes typically 80 press releases per month, or about 4 per working day. One should realise, however, that this is still only a very tiny fraction of all scientific press releases received by journalists. Looking at the European science agency Alphagalileo, astronomy covers only about 10\% of all scientific press releases they distribute. Journalism is really about making a choice and in such conditions, one can be happy if some astronomical news get covered. This is of course not the only place where choices are made. Taking the example of ESO, the European Southern Observatory, one can note that in 2008 there were more than 700 refereed scientific papers published, while there were only 50 press releases, of which only half were based on a scientific paper (the others being organisation news, instrumentation news or press photos). At the sourcce there is thus already a selection by a factor 30!
All in all, the chances that an astronomical scientific paper will be reported upon in the media is less than one in a few hundreds.

Madsen (2001) in his study ``Stars in the Media'', in which he looked at the coverage of astronomy and space science in broadsheet papers in the United Kingdom, France, Spain, Sweden and Denmark, provided some useful conclusions:\\
\begin{itemize}
\item  The choice of topics is influenced by national 
 (cultural, political) aspects, but the narrative 
 (story, rhetoric) is rather uniform;
\item Fundamental research is reported within a 
  narrow scientific frame;
\item  Articles on astrophysics/space currently occupy approximatively
0.1\% of leading European newspapers;
\item There is much more emphasis on health/environment than on astronomy;
\item Science is mostly presented in special sections.
\end{itemize}

\vspace{12pt}

Madsen also emphasised that more effort should be invested to show the role of 
fundamental research for societal development and general culture, and that this may also attract more interest. I can't agree more. Despite the fact that astronomy may be considered humankind's 
boldest attempt to understand the world in which we live, addressing fundamental questions such as ``are we alone?'', ``what is the Universe made of?'', and ``how did it all begin?'', which have deep philosophical, religious, and societal impacts, astronomy is too often limited to the science sections that are accessed by a small audience. We need to bring the message home to the editors that astronomy is not for `geeks' only, but deserves a more prominent place in the media.

\section{Conclusion}

When looking at the presence of astronomy and the media, it is also interesting to have the opinion of the journalists themselves. I have therefore conducted a small survey via e-mail to all journalists subscribed to the ESO media mailing list. This is by no means supposed to be a scientifically accurate survey, but is useful to get a first glance at the `other side'. I submitted to the journalists a series of 5 questions, which are indicated below as well as their answers. Some interesting facts come out.

Most journalists said that they run between 2 and 3 astronomical stories {\it per month}, with some running a few more. Representatives of the online media were generally running more than 5 stories per month. This is due to the fact, as one journalist put it, that  ``online, space is infinite'' and there is not so much struggle with other subjects. This illustrates that given time and space, journalists do find astronomy stories interesting. The majority of journalists said that they have no {\it a priori} about the possible topics to be run, and that the most important when selecting the story is the subject and the availability of a nice image or a video. Some journalists highlighted nevertheless exoplanets and the solar system as the topics they will most likely write about. It seems also that once journalists have made their mind into writing a story, it is not difficult to convince their editor to run it. They also acknowledge, however, that ``they don't make it to the front page'' and are often confined in special sections. And, finally, it is perhaps revealing that almost two-third of the journalists thought that astronomy has the place it deserves in the media. It is also important to note that the journalists said that they won't necessarily increase their coverage of astronomy just because it is the International Year of Astronomy. Good stories 
is what they want and need.

\vspace{12pt}

\begin{table}
\begin{center}
\scriptsize
{\bf Small survey of science journalists}\\
\vspace{12pt}
%\end{center}

1. How often would you run a story related 
to astronomy per month? \\
%\begin{center}
\begin{tabular}{ll}
1 &11\% \\
2-3 &37\% \\ 
4-5 &19\% \\
More than 5 &33\% \\
\end{tabular}
\vspace{12pt}
%\end{center}

2. What are the topics most likely to be run?
(several answers possible) \\
%\begin{center}
\begin{tabular}{lr}
None in particular &55\% \\
Exoplanets &33\% \\
Solar system &29\% \\ 
Cosmology &15\%  \\
Stars and nebulae &11\% \\
Galaxies   &7\% \\
\end{tabular}
\vspace{12pt}
%\end{center}

3. What is most important when selecting the story? 
(several answers possible)\\
%\begin{center}
\begin{tabular}{lr}
Subject & 92\% \\
Availability of a nice image or video  & 74\% \\
Nationalities of the scientists involved  & 22\% \\ 
Names and host institutions of the scientists involved   & 9\% \\
Institution issuing the press release  &  0\% \\
\end{tabular}
\vspace{12pt}
%\end{center}

4. Do you find it difficult to run an 
astronomical story past the chief editor? \\
%\begin{center}
\begin{tabular}{lc}
NO & 80\% \\
YES & 20\% \\
\end{tabular}
\vspace{12pt}
%\end{center}

5. Do you think astronomy has the place it 
deserves in the media? \\
%\begin{center}
\begin{tabular}{lc}
YES & 62\% \\
NO &  38\% \\
\end{tabular}
\vspace{12pt}

\end{center}
\end{table}

It is therefore clear that journalists  appear to be keen to cover astronomy in the media and that most major breakthroughs are covered. ESO estimates a yearly readership in newspapers and magazines of tens of million people worldwide, while it appeared in hundreds of TV news reports or documentaries, potentially reaching hundreds of millions of viewers. The impact is undeniable. This shouldn't hide the fact that more efforts should be done for astronomy to be dealt outside of the special science sections, taking into account its important societal and cultural aspects.

\end{document}